\documentclass[pdftex,twocolumn,epjc3]{svjour3}          

\usepackage{amsmath}

\usepackage{booktabs}
\usepackage{siunitx}

\RequirePackage[T1]{fontenc}

\smartqed  

\RequirePackage{graphicx}
\RequirePackage{mathptmx}      
\RequirePackage{flushend}
\RequirePackage[numbers,sort&compress]{natbib}
\RequirePackage[colorlinks,citecolor=blue,urlcolor=blue,linkcolor=blue]{hyperref}

\journalname{Eur. Phys. J. C}

\begin{document}

\title{Pion chemical potentials as seen from transverse momentum distributions}

\author{
Valeria Zelina Reyna Ortiz\thanksref{e3,addr2}
\and
Maciej Rybczy\'nski\thanksref{e1,addr1}
\and
Zbigniew W\l odarczyk\thanksref{e2,addr1}
}

\thankstext{e3}{e-mail: val.reyna@cern.ch}
\thankstext{e1}{e-mail: Maciej.Rybczynski@ujk.edu.pl}
\thankstext{e2}{e-mail: Zbigniew.Wlodarczyk@ujk.edu.pl}

\institute{
Faculty of Physics, Warsaw University of Technology,
ul. Koszykowa 75, PL-00662 Warsaw, Poland
\label{addr2}
\and
Institute of Physics, Jan Kochanowski University,
ul. Uniwersytecka 7, PL-25406 Kielce, Poland
\label{addr1}
}

\date{Received: date / Accepted: date}

\maketitle

\begin{abstract}
We examine the transverse momentum spectra of negatively charged pions
emitted in proton-proton, proton-antiproton, and nucleus-nucleus collisions 
at beam momenta  from  tens 
to  hundreds of GeV$/c$. The measured spectra are described within 
the framework of nonextensive Tsallis statistics, allowing the corresponding
chemical potentials to be determined through a quantitative comparison
with experimental data. The extracted thermodynamic parameters are
subsequently employed to explore the structure of the phase diagram
in the chemical potential-temperature ($\mu$-$T$) plane, with
particular emphasis on identifying possible signatures of a phase
transition. Within this interpretation, pion production in the
midrapidity region is associated with a phase characterized by
 broken chiral symmetry,  whereas increasing rapidity corresponds to an evolution toward a regime in which chiral symmetry
is restored.

\end{abstract}

\section{Introduction}
\label{Sec:Intro}

The chemical potential  quantifies
the variation of the thermodynamic free energy accompanying a change in
the pion population. Its magnitude provides valuable information about
departures from chemical equilibrium and the possible emergence of
Bose-Einstein condensation in strongly interacting matter. During the
initial evolution of relativistic  collisions, intensive particle
production may generate pion abundances exceeding their equilibrium
values \cite{ZPC58}. As the system subsequently expands and cools,
the approximately conserved pion population can be accommodated through
the development of a finite, temperature-dependent chemical potential
$\mu_{\pi}$. Appreciable positive values of this parameter favor an
enhanced occupation of low-momentum states, leading to an increased
yield of soft pions, particularly negatively charged ones, in the
measured momentum distributions \cite{PRC44}.
When the magnitude of the isospin chemical potential $\mu_I$ reaches
the pion mass, $|\mu_I|\geq m_{\pi}$, the accumulation of charged
pions in the lowest available momentum state becomes energetically
favorable \cite{NPB152}. This condition marks the onset of a quantum
phase transition associated with pion Bose-Einstein condensation,
giving rise to a superfluid-like state of QCD matter \cite{PRC103}.
Investigations of this regime are relevant not only to the phase
structure of strongly interacting systems but also to the description
of astrophysical and cosmological environments, including neutron star
interiors and the cooling evolution of the primordial universe
\cite{PRC103, PRD113}. Furthermore, finite pion chemical potentials
influence the dissipative properties of a hadronic medium by modifying
its shear and bulk viscosities, thereby affecting the collective
expansion of the collision-generated fireball. They may also increase
the contributions from particular hadronic decay processes and alter
dilepton emission under chemically nonequilibrated conditions.
Such modifications provide potentially useful observables for
constraining the dynamical evolution of the strongly interacting
system \cite{PRD113}.

High-energy collisions between heavy nuclei offer an experimental
means of investigating the properties of nuclear matter under
conditions inaccessible in ordinary environments, including the
possible formation of novel phases of strongly interacting matter
\cite{STAR2005}. The reaction initially produces a highly
nonequilibrated, rapidly evolving medium whose characteristic
lifetime is limited to only a few fm/$c$. Its transient nature
precludes straightforward direct observation of the intermediate
stages. Information about the properties of this medium must
therefore be reconstructed from experimentally accessible quantities,
most notably the momentum space distributions and multiparticle
correlations of hadrons emerging after hadronization and freeze-out.
A quantitative interpretation of these measurements requires
knowledge of how the surrounding strongly interacting environment
modifies the characteristics of individual particles. In particular,
temperature-induced changes in mesonic masses can appreciably affect
the resulting hadronic spectra and particle abundances
\cite{Cleymans:2006, Andronic:2009}. Reliable descriptions of
in-medium particle properties are thus indispensable for extracting
the underlying physics from collision observables.

Among the hadronic species produced in nuclear collisions, pions
occupy a distinguished position because of their exceptionally low
masses and their interpretation as pseudo-Goldstone bosons of
spontaneously broken chiral symmetry. These characteristics establish
a close connection between pion dynamics and the restoration of
chiral symmetry in hot and dense QCD matter, making pion-related
observables promising tools for investigating the corresponding
phase transition \cite{Dumitru1995}. Additional motivation comes
from measurements at the Large Hadron Collider, where evidence
for a non-negligible coherent component in charged pion production
has been reported \cite{ALICE2014}. Such observations have been
interpreted as a possible indication of a superfluid state
associated with charged-pion condensation
\cite{Begun2014, Begun2015}. Moreover, the relative production
of negatively and positively charged pions contains information
about the nuclear symmetry energy, an essential ingredient in
the description of  neutron star properties
\cite{Sako}. Thus, the proporties of pions in the medium has attracted a lot of attention.

\section{Analysis of transverse momentum distributions}
\label{Sec:Analysis}

The transverse-momentum ($p_T$) spectra of identified particles constitute
one of the principal experimental observables for investigating the
mechanisms governing hadron production and the dynamical evolution of
strongly interacting systems. In this context, proton-proton collisions
provide an essential reference for interpreting the more complex
phenomena encountered in nuclear interactions
\cite{Becattini:1997rv}. A particularly effective phenomenological
description of particle spectra can be formulated within the
nonextensive statistical approach introduced by Tsallis
\cite{Tsallis:1987eu,Cleymans:2011in,Cleymans:2012ya}.
In this formulation, the single-particle momentum distribution
takes the form
\begin{align}
\frac{d^{3}N}{dp^{3}}
&=
\frac{gV}{(2\pi)^{3}}
\left[
1+(q-1)\frac{E-\mu}{T}
\right]^{\frac{q}{1-q}}
\xrightarrow{q\rightarrow 1}
\nonumber\\
&\frac{gV}{(2\pi)^{3}}
\exp\left(-\frac{E-\mu}{T}\right),
\label{eq:ptdist}
\end{align}
where $T$ denotes the temperature, $\mu$ is the chemical potential,
$V$ represents the effective volume, and $g$  is the degeneracy factor.
The nonextensivity index $q$ controls the departure from conventional
Boltzmann-Gibbs statistics, whose characteristic exponential
dependence is recovered in the limit $q\rightarrow 1$.
Accordingly, the Tsallis formalism provides a generalized
statistical description in which deviations from the standard
equilibrium distribution are incorporated through an additional
parameter.

The physical interpretation of $q$ is closely connected with
statistical fluctuations and other mechanisms responsible for
non-Boltzmann behavior. In particular, Ref.~\cite{Wilk:1999dr}
established the relation
$q-1=\mathrm{Var}(T)/\langle T\rangle^{2}$,
which associates the nonextensivity parameter with the relative
variance of temperature fluctuations. Nevertheless, fluctuating
thermal conditions do not constitute the only possible origin of
Tsallis-like distributions. Similar functional forms may arise
from several distinct dynamical scenarios, as discussed in
Ref.~\cite{Wilk:2012zn}. The applicability of this generalized
statistical framework to a broad class of multiparticle production
phenomena has been demonstrated in numerous investigations;
comprehensive discussions can be found in
Refs.~\cite{Wilk:2012zn,Wilk:2008ue}.

For comparisons with experimentally measured spectra, it is
convenient to express the distribution in terms of transverse
momentum $p_T$, rapidity $y$, and transverse mass
$m_T=\sqrt{m^{2}+p_T^{2}}$. Introducing these kinematic variables
into Eq.~(\ref{eq:ptdist}) yields
\begin{align}
\frac{d^{2}N}{p_Tdp_Tdy}
&=
gV\frac{m_T\cosh(y)}{(2\pi)^{2}}
\nonumber\\
&\quad\times
\left[
1+(q-1)
\frac{m_T\cosh(y)-\mu}{T}
\right]^{\frac{q}{1-q}}.
\label{eq:ptdist2}
\end{align}
The ability of this parametrization to reproduce transverse momentum
distributions has been extensively investigated over a wide range
of collision energies. Successful descriptions of spectra measured
in proton-proton interactions have been reported at RHIC energies
of $\sqrt{s}=62.4$ and $200$~GeV, as well as at the LHC for
$\sqrt{s}=0.9$, $2.76$, and $7.0$~TeV
\cite{Cleymans:2011in,Sena:2012ds,Adare:2010fe,
Khandai:2013gva,Wong:2012zr}. The energy dependence of the
spectral shapes was examined further in
Ref.~\cite{Cleymans:2013rfq}, where measurements at
$\sqrt{s}=0.54$, $0.9$, $2.36$, and $7$~TeV were analyzed
using Eq.~(\ref{eq:ptdist2}). This approach was subsequently
extended in Ref.~\cite{RW:2014} to substantially lower
center-of-mass energies, namely
$\sqrt{s}=6.27$, $7.74$, $8.76$, $12.32$, and $17.3$~GeV,
employing proton-proton data collected by the NA61/SHINE
experiment \cite{NA61SHINE:2013tiv}.

The NA61/SHINE measurements offer two important advantages for
such studies. Besides extending the accessible energy domain
toward lower incident momenta, they provide information on the
spectral behavior at small transverse momenta. This region is
particularly relevant because the extracted temperature and
effective volume exhibit considerable sensitivity to the
low $p_T$ contribution. Consequently, incorporating measurements
closer to the origin of the transverse momentum spectrum can
improve the determination and interpretation of these parameters.

In the present analysis, the measured distributions of negatively
charged pions are represented by the Tsallis distribution.
The initial
parametrization, with the degeneracy factor fixed at $g_{\pi^-}=1$,  is performed under the assumption of a vanishing
chemical potential, $\mu=0$, leading to the reference parameters
$T_0$ and $V_0$. Under these conditions, the fitting function
reduces to
\begin{align}
\frac{d^{2}N}{p_Tdp_Tdy}
&=
gV_0\frac{m_T\cosh(y)}{(2\pi)^{2}}
\left[
1+(q-1)
\frac{m_T\cosh(y)}{T_0}
\right]^{\frac{q}{1-q}}.
\label{eq:ptdist0}
\end{align}

It is important to recognize that, for a specified value of the
nonextensivity index $q$, the temperature and normalization
volume are not independent of the chemical potential. Instead,
the corresponding parameters are related through the
reparametrization
\begin{equation}
T=T_0+(q-1)\mu,
\label{eq:EqT}
\end{equation}
and
\begin{equation}
V=
V_0\left[
1-(q-1)\frac{\mu}{T_0}
\right]^{\frac{q}{1-q}}
=
V_0\left(\frac{T}{T_0}\right)^{\frac{q}{1-q}}.
\label{eq:EqV}
\end{equation}
Thus, once the reference quantities $T_0$ and $V_0$ have been
determined at $\mu=0$, their values corresponding to a finite
chemical potential can be obtained for fixed $q$ using the
above relations \cite{Cleymans:2013rfq}.

The phenomenological investigation encompasses several collision
systems, including proton-proton ($p+p$), proton-antiproton
($p+\bar{p}$), and nucleus-nucleus ($A+A$) interactions.
For proton-proton collisions, the experimental input consists
of measurements performed by NA61/SHINE at the CERN Super
Proton Synchrotron and by ATLAS at the Large Hadron Collider
\cite{NA61SHINE:2013tiv,ATLAS:2010jvh}. The proton-antiproton
sample is based on results reported by the UA1 experiment
at the CERN SPS \cite{UA1:1989bou}. The nuclear collision
data include $Be+Be$ measurements from NA61/SHINE
\cite{NA61SHINE:2020ggt}, $Cu+Cu$ results obtained by
PHOBOS \cite{PHOBOS:2005jpg}, and $Au+Au$ spectra measured
by STAR \cite{STAR:2017sal}, with the latter two experiments
operating at RHIC at Brookhaven National Laboratory.

Most of the selected measurements correspond to the midrapidity
region. Two datasets require special consideration because of
their different kinematic acceptances. The $Be+Be$ spectra
were recorded within $0<y<0.2$, whereas the $Cu+Cu$
measurements cover the pseudorapidity range
$0.2<\eta<1.4$, where $\eta$ denotes pseudorapidity.
To account for the displacement of these intervals from
midrapidity, the fitted reference temperature was adjusted
according to
$T_0\longrightarrow T_0/\cosh(y_{\mathrm{mid}})$,
where the central values $y_{\mathrm{mid}}=0.1$ and
$y_{\mathrm{mid}}=0.8$ were adopted for $Be+Be$ and
$Cu+Cu$ collisions, respectively. The resulting
nonextensivity indices $q$, reference temperatures $T_0$,
and effective radii $R_0$, together with the remaining
parameters characterizing the description of negatively
charged pion spectra, are summarized in
Table~\ref{tab:fit_parameters}.

\begin{table*}[t]
\centering
\caption{Parameters characterizing the transverse-momentum
spectra of negatively charged pions for the investigated
collision systems, center-of-mass energies, and centrality
classes.}
\label{tab:fit_parameters}

\small
\setlength{\tabcolsep}{3.5pt}
\renewcommand{\arraystretch}{1.15}

\begin{tabular*}{\textwidth}{
@{\extracolsep{\fill}}
l
c
S[table-format=4.1]
c
S[table-format=1.4]
S[table-format=1.4]
S[table-format=2.2]
S[table-format=1.3]
S[table-format=1.6]
S[table-format=1.6]
@{}
}
\toprule
{System}
& {Centrality}
& {$\sqrt{s_{NN}}$}
& {Ref.}
& {$q$}
& {$T_{0}$}
& {$R_{0}$}
& {$\kappa$}
& {$T$}
& {$\mu$}
\\[-1mm]

&
&
{(GeV)}
&
&
&
{(GeV)}
&
{(fm)}
&
&
{(GeV)}
&
{(GeV)}
\\
\midrule

$p+p$
& --
& 6.3
& {\cite{NA61SHINE:2013tiv}}
& 1.05
& 0.0945
& 1.43
& 1.9
& 0.103
& 0.181
\\

$p+p$
& --
& 7.7
& {\cite{NA61SHINE:2013tiv}}
& 1.053
& 0.0945
& 1.49
& 1.9
& 0.104
& 0.181
\\

$p+p$
& --
& 8.8
& {\cite{NA61SHINE:2013tiv}}
& 1.056
& 0.0945
& 1.51
& 1.9
& 0.105
& 0.181
\\

$p+p$
& --
& 12.3
& {\cite{NA61SHINE:2013tiv}}
& 1.059
& 0.0975
& 1.53
& 1.9
& 0.108
& 0.187
\\

$p+p$
& --
& 17.3
& {\cite{NA61SHINE:2013tiv}}
& 1.073
& 0.0925
& 1.65
& 1.9
& 0.105
& 0.177
\\

$p+p$
& --
& 900.0
& {\cite{ATLAS:2010jvh}}
& 1.125
& 0.099
& 1.93
& 1.9
& 0.123
& 0.189
\\

$p+p$
& --
& 2360.0
& {\cite{ATLAS:2010jvh}}
& 1.14
& 0.097
& 2.08
& 1.9
& 0.123
& 0.185
\\

$p+p$
& --
& 7000.0
& {\cite{ATLAS:2010jvh}}
& 1.151
& 0.097
& 2.33
& 1.9
& 0.125
& 0.185
\\

$p+\bar{p}$
& --
& 200.0
& {\cite{UA1:1989bou}}
& 1.1
& 0.099
& 1.93
& 1.9
& 0.118
& 0.189
\\

$p+\bar{p}$
& --
& 500.0
& {\cite{UA1:1989bou}}
& 1.112
& 0.099
& 1.99
& 1.9
& 0.120
& 0.189
\\

$p+\bar{p}$
& --
& 900.0
& {\cite{UA1:1989bou}}
& 1.119
& 0.099
& 2.11
& 1.9
& 0.121
& 0.189
\\

$Be+Be$
& 0--20\%
& 17.0
& {\cite{NA61SHINE:2020ggt}}
& 1.11
& 0.0796
& 3.48
& 1.6
& 0.091
& 0.108
\\

$Cu+Cu$
& 0--6\%
& 62.4
& {\cite{PHOBOS:2005jpg}}
& 1.06
& 0.138
& 4.43
& 1.25
& 0.143
& 0.089
\\

$Au+Au$
& 0--5\%
& 7.7
& {\cite{STAR:2017sal}}
& 1.06
& 0.108
& 7.36
& 1.7
& 0.118
& 0.170
\\

$Au+Au$
& 0--5\%
& 11.5
& {\cite{STAR:2017sal}}
& 1.065
& 0.108
& 7.97
& 1.7
& 0.119
& 0.169
\\

$Au+Au$
& 0--5\%
& 19.6
& {\cite{STAR:2017sal}}
& 1.073
& 0.108
& 8.67
& 1.7
& 0.120
& 0.169
\\

$Au+Au$
& 0--5\%
& 27.0
& {\cite{STAR:2017sal}}
& 1.077
& 0.108
& 8.83
& 1.7
& 0.121
& 0.169
\\

$Au+Au$
& 0--5\%
& 39.0
& {\cite{STAR:2017sal}}
& 1.082
& 0.108
& 8.92
& 1.7
& 0.122
& 0.169
\\

$Au+Au$
& 30--40\%
& 7.7
& {\cite{STAR:2017sal}}
& 1.06
& 0.104
& 5.07
& 1.5
& 0.111
& 0.123
\\

$Au+Au$
& 30--40\%
& 11.5
& {\cite{STAR:2017sal}}
& 1.065
& 0.104
& 5.34
& 1.5
& 0.112
& 0.123
\\

$Au+Au$
& 30--40\%
& 19.6
& {\cite{STAR:2017sal}}
& 1.076
& 0.104
& 5.8
& 1.5
& 0.113
& 0.123
\\

$Au+Au$
& 30--40\%
& 27.0
& {\cite{STAR:2017sal}}
& 1.081
& 0.104
& 6.01
& 1.5
& 0.114
& 0.122
\\

$Au+Au$
& 30--40\%
& 39.0
& {\cite{STAR:2017sal}}
& 1.086
& 0.104
& 6.08
& 1.5
& 0.114
& 0.122
\\

$Au+Au$
& 70--80\%
& 7.7
& {\cite{STAR:2017sal}}
& 1.06
& 0.09
& 2.91
& 1.33
& 0.094
& 0.074
\\

$Au+Au$
& 70--80\%
& 11.5
& {\cite{STAR:2017sal}}
& 1.07
& 0.09
& 3.07
& 1.33
& 0.095
& 0.074
\\

$Au+Au$
& 70--80\%
& 19.6
& {\cite{STAR:2017sal}}
& 1.078
& 0.09
& 3.31
& 1.33
& 0.096
& 0.074
\\

$Au+Au$
& 70--80\%
& 27.0
& {\cite{STAR:2017sal}}
& 1.084
& 0.09
& 3.35
& 1.33
& 0.096
& 0.073
\\

$Au+Au$
& 70--80\%
& 39.0
& {\cite{STAR:2017sal}}
& 1.097
& 0.09
& 3.42
& 1.33
& 0.097
& 0.073
\\

\bottomrule
\end{tabular*}
\end{table*}

\section{The pion chemical potential}
\label{Sec:EnDep}

Although the parameter $V_0$ enters the normalization of the
transverse-momentum distribution in Eq.~(\ref{eq:ptdist2}),
its interpretation as a physical volume requires some caution.
In particular, the effective volume extracted from the Tsallis
parametrization cannot be directly identified with the spatial
extent of the particle-emitting source inferred from
Hanbury Brown--Twiss (HBT) interferometry
\cite{Aamodt:2010jj,Aggarwal:2010aa}.

A more physically motivated estimate of the characteristic
source dimensions can nevertheless be obtained by relating
the fitted radius to experimentally determined HBT scales.
The reference radius $R_0$ corresponds to the parametrization
with vanishing chemical potential. According to
Eqs.~(\ref{eq:EqT}) and~(\ref{eq:EqV}), introducing a
positive chemical potential leads to a reduction of the
corresponding effective volume and, consequently, of the
associated source radius. Comparison with the HBT
measurements suggests that the relevant spatial scale
can be approximated by
$R_{\mathrm{HBT}}\simeq R=R_0/\kappa$,
where $\kappa$ specifies the rescaling between the
reference radius and the experimentally constrained
source dimension.
This correspondence motivates the following relation
between the effective volumes evaluated at zero and
finite chemical potentials:
\begin{equation}
V_{0}=V_{\mu}\,\kappa^{3}.
\label{eq:Vmu}
\end{equation}

Combining this prescription with the temperature and
volume transformations given by
Eqs.~(\ref{eq:EqT}) and~(\ref{eq:EqV}) provides an
explicit expression for the pion chemical potential
in terms of the reference temperature, the
nonextensivity parameter, and the geometrical
rescaling factor:
\begin{equation}
\mu=
\frac{T_0}{q-1}
\left(
\kappa^{3(q-1)/q}-1
\right).
\label{eq:mukappa}
\end{equation}

\begin{figure}
\begin{minipage}{\columnwidth}
\centering
\centerline{
\includegraphics[width=0.97\columnwidth]{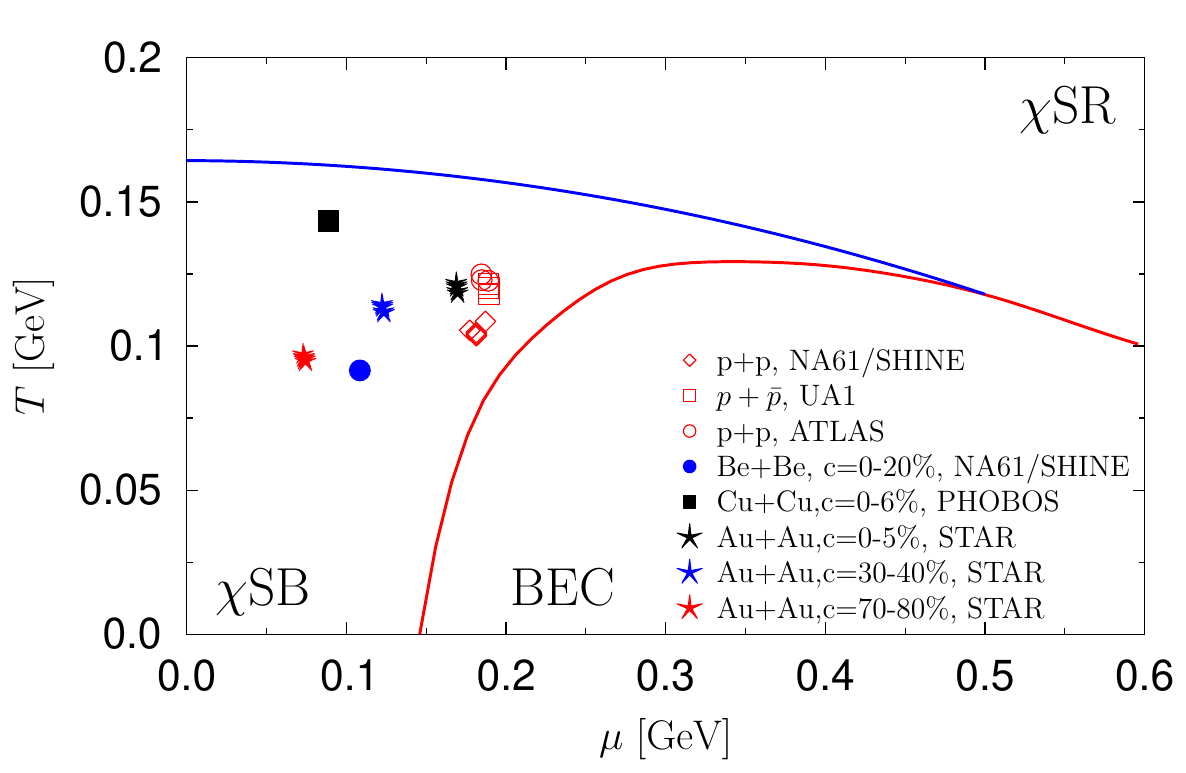}
}
\end{minipage}
\caption{\small (Color online) Temperatures
and pion chemical potentials extracted for the
investigated collision systems in midrapidity region, displayed against
the phase structure associated with chiral symmetry crossover
and pion Bose-Einstein condensation in the
$T$--$\mu$ plane. The blue line distinguishes
the chirally symmetry broken ($\chi$SB) and chiral symmetry
restored ($\chi$SR) phases, whereas the red
boundary identifies the domain of pion
condensation (BEC)
\cite{XuanminPRD,XuanminChPC}.}
\label{fig:mu_t}
\end{figure}

The resulting thermodynamic parameters permit a
systematic examination of the freeze-out conditions
in the temperature - chemical potential ($T$--$\mu$)
plane. Figure~\ref{fig:mu_t} presents the values
obtained for the investigated collision systems,
together with the theoretical phase boundaries
associated with chiral symmetry restoration and
pion condensation. The experimental points
correspond to negatively charged pions measured
in the midrapidity region. Independently of the
colliding species and, different centralities of nucleus-nucleus collisions,
 these points are located
within the chiral symmetry broken phase ($\chi$SB).

\begin{figure}
\begin{minipage}{\columnwidth}
\centering
\centerline{
\includegraphics[width=0.97\columnwidth]{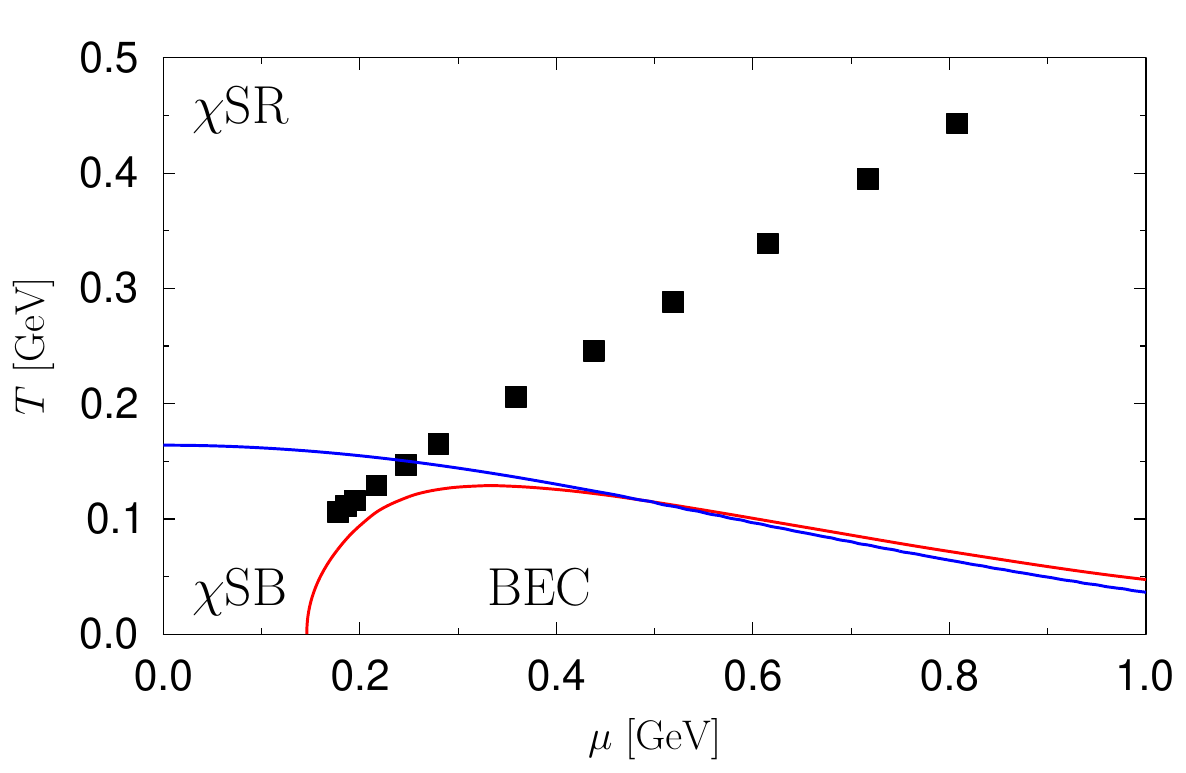}
}
\end{minipage}
\caption{\small (Color online) Rapidity dependence
of the temperature and pion chemical potential
determined from negatively charged pion spectra
in proton-proton interactions at
$\sqrt{s}=17.3$~GeV, measured by NA61/SHINE
\cite{NA61SHINE:2013tiv}. The results correspond
to successive rapidity bins
$0.2k<y<0.2(k+1)$, where
$k=0,\ldots,11$ increases from bottom to top.
The blue curve indicates the separation
between the chiral symmetry broken ($\chi$SB)
and restored ($\chi$SR) phases, while the
red curve delineates the region associated
with pion Bose-Einstein condensation
\cite{XuanminPRD,XuanminChPC}.}
\label{fig:mu_t_y}
\end{figure}

The rapidity dependence reveals an additional
characteristic feature of the extracted thermodynamic
quantities. Both the effective temperature and the
pion chemical potential exhibit a systematic increase
toward larger rapidities, approximately following
a hyperbolic-cosine dependence (approximately as: 
$\mu\simeq 0.0196+0.166\cosh(y)$,
and
$T\simeq 0.024+0.088\cosh(y)$).

This behavior is illustrated in
Fig.~\ref{fig:mu_t_y}, where the extracted
temperature and chemical potential are presented
for negatively charged pions produced in
proton-proton collisions at
$\sqrt{s}=17.3$~GeV. The analysis employs
the rapidity intervals
$0.2k<y<0.2(k+1)$, with
$k=0,\ldots,11$, using measurements reported
by the NA61/SHINE collaboration
\cite{NA61SHINE:2013tiv}.

The displacement of the extracted thermodynamic
parameters with increasing rapidity is particularly
significant when interpreted in the context of
the QCD phase diagram. Whereas pion production
near midrapidity is associated with the
chirally broken regime, the evolution toward
larger rapidities moves the corresponding
points across the chiral phase boundary.
Within the adopted phenomenological description,
this behavior indicates a transition from
the  broken chiral symmetry
phase ($\chi$SB) to the chiral symmetry restored
phase ($\chi$SR).

\section{Comparison with chiral phase transition scenario}
\label{Sec:DifPar}

The realization of chiral symmetry in quantum chromodynamics
(QCD) provides a fundamental framework for understanding
the properties of strongly interacting matter under
different thermodynamic conditions. In the limit of
vanishing current quark masses, the QCD Lagrangian
possesses a symmetry relating independently transforming
left- and right-handed quark fields. This symmetry,
however, is not realized in the same manner in the
physical vacuum, where the formation of a nonvanishing
quark condensate, $\langle\bar{q}q\rangle$, leads to
its spontaneous breaking. The resulting dynamical
generation of constituent quark masses constitutes
an essential aspect of nonperturbative QCD and is
closely connected with the origin of hadronic masses,
which account for the overwhelming majority of the
mass of ordinary visible matter. An additional
consequence is the appearance of light pseudoscalar
mesons, with pions representing the pseudo-Goldstone
modes associated with the spontaneously broken
symmetry.

The thermodynamic evolution of strongly interacting
matter may considerably modify this vacuum structure.
At sufficiently elevated temperatures or densities,
the magnitude of the chiral condensate decreases,
accompanied by a reduction in the dynamically
generated contribution to the quark masses.
This behavior characterizes the approach to the
chiral symmetry restored regime ($\chi$SR), in contrast
to the low-temperature phase exhibiting spontaneous
chiral symmetry breaking ($\chi$SB). At small
baryochemical potential, the characteristic
temperature associated with the chiral crossover
is approximately $T_c\simeq150$-$160$~MeV.
Such conditions are expected to have occurred
during the early evolution of the Universe and
can be approached experimentally in relativistic
heavy-ion collisions. Other extreme environments,
including those involving sufficiently strong
magnetic fields or acceleration, may also influence
the realization of chiral symmetry.

Because pions emerge from the nonperturbative
dynamics underlying spontaneous chiral symmetry
breaking, their behavior in a strongly interacting
medium cannot generally be addressed through
ordinary perturbative methods. Investigations
of the corresponding phase structure therefore
rely on theoretical approaches capable of
incorporating essential nonperturbative effects.
These include lattice QCD calculations
\cite{Ishii2017,Brandt2015}, treatments based
on Dyson-Schwinger equations \cite{Fisher2019},
functional renormalization-group techniques
\cite{Wang2017}, and effective descriptions
constructed within the Nambu-Jona-Lasinio
framework \cite{Xia2013,Xia2019}. Among
alternative formulations, holographic hard-wall
and soft-wall models offer useful descriptions
of hadronic spectroscopy and provide a basis
for examining the thermodynamic evolution of
chiral symmetry \cite{XuanminPRD,XuanminChPC}.

The QCD phase structure becomes particularly
rich in the presence of a finite isospin
chemical potential. An imbalance between
up- and down-quark populations, which is
closely related to the isospin asymmetry of
nuclear matter, may favor the formation
of a charged-pion condensate
\cite{Carignano2017}. In this state,
macroscopic occupation of the lowest-energy
mode by charged pions, either $\pi^+$ or
$\pi^-$ depending on the sign of the
isospin chemical potential, gives rise
to Bose-Einstein condensation and the
spontaneous breaking of the associated
global U(1) symmetry. At zero temperature,
the onset of this phase is determined
by a threshold involving the pion mass.
Depending on the normalization adopted
for the isospin chemical potential,
the critical condition is conventionally
expressed as $\mu_I=m_\pi/2$ or
$\mu_I=m_\pi$.

\begin{figure}
\begin{minipage}{\columnwidth}
\centering
\centerline{
\includegraphics[width=0.97\columnwidth]{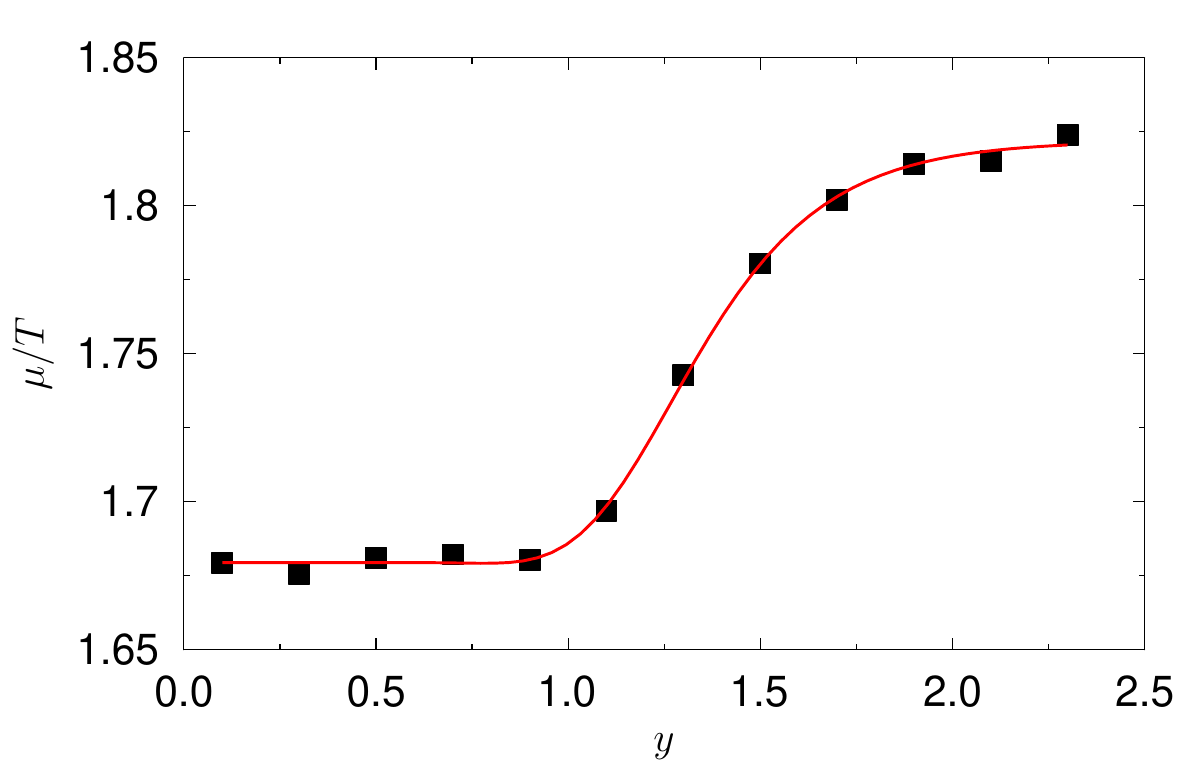}
}
\end{minipage}
\caption{\small (Color online) Variation of
the pion chemical potential to temperature
ratio, $\mu/T$, with rapidity in
proton-proton collisions at
$\sqrt{s}=17.3$~GeV. The experimental
results correspond to negatively charged
pions measured by the NA61/SHINE
collaboration \cite{NA61SHINE:2013tiv}.
The curve represents the Gompertz
parametrization defined by
Eq.~(\ref{eq:Gomparetz}).}
\label{fig:mu_t_ratio}
\end{figure}

The thermodynamic parameters reconstructed
from the transverse momentum spectra can be
examined in relation to these theoretically
predicted phases. In particular, the results
displayed in Fig.~\ref{fig:mu_t_y} indicate
that the rapidity evolution of the extracted
temperature and pion chemical potential is
accompanied by a pronounced variation in
their ratio, $\mu/T$. This variation becomes
especially apparent in the region associated
with the transition between the chirally
broken and chirally restored regimes.
To quantify the observed behavior, the
rapidity dependence of $\mu/T$ is represented
by a Gompertz-type function,
\begin{equation}
\frac{\mu}{T}
=
a+b\exp\left[
-\exp\left(-c(y-y_c)\right)
\right],
\label{eq:Gomparetz}
\end{equation}
with the parameters: 
$a=1.68$, $b=0.143$, $c=4.44$, and
$y_c=1.26$. The corresponding description
is illustrated in Fig.~\ref{fig:mu_t_ratio},
which shows the extracted $\mu/T$ values
for negatively charged pions produced
in proton-proton interactions at
$\sqrt{s}=17.3$~GeV, using the
NA61/SHINE measurements
\cite{NA61SHINE:2013tiv}.

The modification of the pion chemical
potential relative to the temperature
may be considered in the context of
changes in the hadronic excitation
spectrum accompanying chiral symmetry
restoration. In the chirally restored
regime, the scalar $\sigma$ meson
becomes the chiral partner of the pion,
while the spectral properties of
vector and axial-vector mesons approach
degeneracy. Such modifications of
the hadronic degrees of freedom may
influence particle production during
the subsequent evolution of the system.
In particular, chiral symmetry
restoration has been proposed as
a possible mechanism contributing
to an enhanced pion abundance
at freeze-out \cite{Song}.

\section{Discussion and conclusions}
\label{Sec:DisCon}

The realization of chiral symmetry constitutes an essential aspect
of the theoretical description of strong interactions and provides
fundamental insight into the nonperturbative structure of quantum
chromodynamics (QCD) \cite{Nefediev,Meisner}. In particular,
the phenomenology of strongly interacting matter at low energies
is intimately connected with the spontaneous breaking of chiral
symmetry in the QCD vacuum. Among the consequences of this mechanism
is the emergence of exceptionally light pseudoscalar mesons,
identified with the pseudo-Goldstone modes of the broken symmetry.
Pions therefore occupy a distinguished position in low-energy
hadron dynamics, where they constitute the fundamental degrees
of freedom of chiral perturbation theory, the effective field
theoretical description of QCD in this regime. Conversely,
the possible restoration of chiral symmetry under extreme
thermodynamic conditions represents an important aspect of
the QCD phase structure, with implications extending to the
evolution of the early Universe and the properties of dense
astrophysical objects, including neutron stars.

Of particular interest is the connection between 
the pseudo-Goldstone character of pions and the different
realizations of chiral symmetry encountered across distinct
physical regimes. At sufficiently low temperatures and
energies, the nonvanishing quark condensate characterizes
the spontaneously broken phase. Increasing the temperature
or density of strongly interacting matter may progressively
suppress this condensate, eventually leading toward the
chirally restored regime. In the ultrarelativistic limit,
the distinction between helicity and chirality also becomes
particularly relevant for the underlying fermionic degrees
of freedom. For effectively massless quarks, helicity
coincides with chirality, whereas finite quark masses
couple the left- and right-chiral components. Unlike
chirality, the helicity of a massive particle is not
invariant under arbitrary changes of inertial reference
frame, since an appropriate Lorentz boost can reverse
the relative orientation of its spin and momentum.
These considerations illustrate the importance of
distinguishing the collective restoration of chiral
symmetry in strongly interacting matter from the
kinematic properties of individual ultrarelativistic
particles.

Further theoretical developments, complemented by
precision measurements in relativistic collision
experiments, are expected to provide a more detailed
understanding of the manifestations of chiral symmetry
in QCD. Such investigations should contribute to
clarifying the microscopic dynamics of strongly
interacting matter and the significance of chiral symmetry 
phenomena for high energy particle physics,
nuclear astrophysics, and cosmology.

\begin{acknowledgements}
In preparation of this publication we used the resources of the Center for Computing and Computer Modeling of the Faculty of Natural Sciences of the Jan Kochanowski University in Kielce, modernised from the funds of the Polish Ministry of Science and Higher Education in the “Regional Excellence Initiative” programme under the project RID/SP/00015/2024/01. V.Z. Reyna Ortiz’s work was supported by the project RID/SP/00015/2024/01.
\end{acknowledgements}


\begin{thebibliography}{9}


\bibitem{ZPC58}
D. Bandyopadhyay, M. I. Gorenstein, H. Stöcker, W. Greiner, and H. Sorge,
Zeitschrift für Physik C {\bf 58},  461–464 (1993)

\bibitem{PRC44}
Gorenstein, Mark I. and Yang, Shin Nan,
Phys. Rev. C {\bf 44}, 2875 (1991).

\bibitem{NPB152}
M. Loewe, C. Villavicencio,
Nuclear Physics B {\bf 152} -  Proceedings Supplements, 250-253 ( 2006)


\bibitem{PRC103}
Stashko, O. S. and Savchuk, O. V. and Poberezhnyuk, R. V. and Vovchenko, V. and Gorenstein, M. I.
Phys. Rev. C{ \bf 103}, 065201, (2021)

\bibitem{PRD113}
Bandyopadhyay, Aritra and Islam, Chowdhury Aminul and Redlich, Krzysztof and Sasaki, Chihiro,
Phys. Rev. D { \bf 113}, 114044, (2026).

\bibitem{STAR2005}
 J. Adams et al. (STAR Collaboration), Nucl. Phys. A {\bf 757},
102 (2005).

\bibitem{Cleymans:2006}
J. Cleymans, I. Kraus, H. Oeschler, K. Redlich, and S.
Wheaton, Phys. Rev. C  {\bf 74}, 034903 (2006).

\bibitem{Andronic:2009}
A. Andronic, P. Braun-Munzinger, and J. Stachel, Phys.
Lett. B 673, 142 (2009); 678, 516(E) (2009).

\bibitem{Dumitru1995}
 A. Dumitru, U. Katscher, J. A. Maruhn, H. Stoecker, W.
Greiner, and D. H. Rischke, Phys. Rev. C 51, 2166 (1995).

\bibitem{ALICE2014}
B. B. Abelev et al. (ALICE Collaboration), Phys. Rev. C
89, 024911 (2014).

\bibitem{Begun2014}
V. Begun, W. Florkowski, and M. Rybczynski, Phys. Rev.
C 90, 014906 (2014).

\bibitem{Begun2015}
V. Begun and W. Florkowski, Phys. Rev. C 91, 054909
(2015).

\bibitem{Sako}
 M. Sako, T. Murakami, Y. Nakai, Y. Ichikawa, K. Ieki, S.
Imajo, T. Isobe, M. Matsushita, J. Murata, and S.
Nishimura et al., arXiv:1409.3322.



\bibitem{Becattini:1997rv} 
  F.~Becattini and U.~W.~Heinz,
  Z.\ Phys.\ C {\bf 76}, 269 (1997)
  [Erratum-ibid.\ C {\bf 76}, 578 (1997)]
  [hep-ph/9702274].

\bibitem{Tsallis:1987eu} 
  C.~Tsallis,
  J.\ Statist.\ Phys.\  {\bf 52}, 479 (1988);
  Eur.\ Phys.\ J.\ A {\bf 40}, 257 (2009)
  [arXiv:0812.4370 [physics.data-an]] and
  Introduction to Nonextensive Statistical Mechanics (Springer, 2009). 
  For an updated bibliography on this subject, see \url{http://tsallis.cat.cbpf.br/biblio.htm}.

  
\bibitem{Cleymans:2011in} 
  J.~Cleymans and D.~Worku,
  J.\ Phys.\ G {\bf 39}, 025006 (2012)
  [arXiv:1110.5526 [hep-ph]].

\bibitem{Cleymans:2012ya} 
  J.~Cleymans and D.~Worku,
  Eur.\ Phys.\ J.\ A {\bf 48}, 160 (2012)
  [arXiv:1203.4343 [hep-ph]].

\bibitem{Wilk:1999dr} 
  G.~Wilk and Z.~Wlodarczyk,
  Phys.\ Rev.\ Lett.\  {\bf 84}, 2770 (2000)
  [hep-ph/9908459].

\bibitem{Wilk:2012zn} 
  G.~Wilk and Z.~Wlodarczyk,
  Eur.\ Phys.\ J.\ A {\bf 48}, 161 (2012)
  [arXiv:1203.4452 [hep-ph]].

\bibitem{Wilk:2008ue} 
  G.~Wilk and Z.~Wlodarczyk,
  Eur.\ Phys.\ J.\ A {\bf 40}, 299 (2009)
  [arXiv:0810.2939 [hep-ph]]; 
  Central Eur.\ J.\ Phys.\  {\bf 10}, 568 (2012)
  [arXiv:1110.4220 [hep-ph]].

\bibitem{Sena:2012ds} 
  I.~Sena and A.~Deppman,
  Eur.\ Phys.\ J.\ A {\bf 49}, 17 (2013)
  [arXiv:1209.2367 [hep-ex]].

\bibitem{Adare:2010fe} 
  A.~Adare {\it et al.}  [PHENIX Collaboration],
  Phys.\ Rev.\ D {\bf 83}, 052004 (2011)
  [arXiv:1005.3674 [hep-ex]].

\bibitem{Khandai:2013gva} 
  P.~K.~Khandai, P.~Sett, P.~Shukla and V.~Singh,
  Int.\ J.\ Mod.\ Phys.\ A {\bf 28}, 1350066 (2013)
  [arXiv:1304.6224 [hep-ph]].

\bibitem{Wong:2012zr} 
  C.~-Y.~Wong and G.~Wilk,
  Acta Phys.\ Polon.\ B {\bf 43}, 2047 (2012)
  [arXiv:1210.3661 [hep-ph]].

\bibitem{Cleymans:2013rfq} 
  J.~Cleymans, G.~I.~Lykasov, A.~S.~Parvan, A.~S.~Sorin, O.~V.~Teryaev and D.~Worku,
  Phys.\ Lett.\ B {\bf 723}, 351 (2013)
  [arXiv:1302.1970 [hep-ph]].

\bibitem{RW:2014}
Maciej Rybczy\' nski, Zbigniew Włodarczyk,
Eur. Phys. J. C (2014) 74:2785
DOI 10.1140/epjc/s10052-014-2785-7


\bibitem{NA61SHINE:2013tiv}
N.~Abgrall \textit{et al.} [NA61/SHINE],
Eur. Phys. J. C \textbf{74}, no.3, 2794 (2014)
doi:10.1140/epjc/s10052-014-2794-6
[arXiv:1310.2417 [hep-ex]].

\bibitem{ATLAS:2010jvh}
G.~Aad \textit{et al.} [ATLAS],
New J. Phys. \textbf{13}, 053033 (2011)
doi:10.1088/1367-2630/13/5/053033
[arXiv:1012.5104 [hep-ex]].

\bibitem{UA1:1989bou}
C.~Albajar \textit{et al.} [UA1],
Nucl. Phys. B \textbf{335}, 261-287 (1990)
doi:10.1016/0550-3213(90)90493-W

\bibitem{NA61SHINE:2020ggt}
A.~Acharya \textit{et al.} [NA61/SHINE],
Eur. Phys. J. C \textbf{80}, no.10, 961 (2020)
[erratum: Eur. Phys. J. C \textbf{81}, no.2, 144 (2021)]
doi:10.1140/epjc/s10052-020-08514-6
[arXiv:2008.06277 [nucl-ex]].

\bibitem{PHOBOS:2005jpg}
B.~Alver \textit{et al.} [PHOBOS],
Phys. Rev. Lett. \textbf{96}, 212301 (2006)
doi:10.1103/PhysRevLett.96.212301
[arXiv:nucl-ex/0512016 [nucl-ex]].

\bibitem{STAR:2017sal}
L.~Adamczyk \textit{et al.} [STAR],
Phys. Rev. C \textbf{96}, no.4, 044904 (2017)
doi:10.1103/PhysRevC.96.044904
[arXiv:1701.07065 [nucl-ex]].

\bibitem{Aamodt:2010jj} 
  K.~Aamodt {\it et al.}  [ALICE Collaboration],
  Phys.\ Rev.\ D {\bf 82}, 052001 (2010)
  [arXiv:1007.0516 [hep-ex]].

\bibitem{Aggarwal:2010aa} 
  M.~M.~Aggarwal {\it et al.}  [STAR Collaboration],
  Phys.\ Rev.\ C {\bf 83}, 064905 (2011)
  [arXiv:1004.0925 [nucl-ex]].

\bibitem{XuanminPRD}
Xuanmin Cao,  Hui Liu   and Danning Li 
Phys.  Rev. D 102, 126014 (2020)
DOI: 10.1103/PhysRevD.102.126014

\bibitem{XuanminChPC}
Xuanmin Cao,  Hui Liu,  Danning Li,  Guanning Ou
Chinese Physics C Vol. 44, No. 8 (2020) 083106
DOI: 10.1088/1674-1137/44/8/083106

\bibitem{Ishii2017}
M. Ishii, H. Kouno, and M. Yahiro, Phys. Rev. D 95,114022 (2017).

\bibitem{Brandt2015}
 B. B. Brandt, A. Francis, H. B. Meyer, and D. Robaina, Phys. Rev. D 92, 094510 (2015).

\bibitem{Fisher2019}
C. S. Fischer, Prog. Part. Nucl. Phys. 105, 1 (2019).

\bibitem{Wang2017}
Z. Wang and P. Zhuang, Phys. Rev. D 96, 014006 (2017).


\bibitem{Xia2013}
T. Xia, L. He, and P. Zhuang, Phys. Rev. D 88, 056013 (2013).

\bibitem{Xia2019}
 T. Xia, J. Hu, and S. Mao, Chin. Phys. C 43, 054103 (2019).


\bibitem{Carignano2017}
S. Carignano, L. Lepori, A. Mammarella, M. Mannarelli, G, Pagliaroli,
Eur. Phys. J. A  53, 35 (2017)

\bibitem{Song}
Ch. Song, V. Koch,
Physics Letters B 404 (1997) 1

\bibitem{Nefediev}
A. Nefediev, Chiral symmetry and its breaking,
Editor(s): Ch. S. Fischer,
Encyclopedia of Particle Physics (First Edition), Elsevier, 2026, Pages 93-106,
https://doi.org/10.1016/B978-0-443-26598-3.00086-9.

\bibitem{Meisner}
U-G. Meißner, Chiral perturbation theory,
Editor(s): Ch. S. Fischer,
Encyclopedia of Particle Physics (First Edition), Elsevier, 2026, Pages 485-508,,
https://doi.org/10.1016/B978-0-443-26598-3.00008-0.



\end{thebibliography}
\end{document}